\documentclass[prd,aps,showpacs,preprintnumbers,amssymb]{revtex4}

\usepackage{color}
\usepackage{epsf}

\def\e3p{$\eta \rightarrow 3 \pi$}

\begin{document}
\title{%
\hfill{\normalsize\vbox{%
\hbox{}
 }}\\
{Higgs boson mass from gauge invariant operators}}

\author{Renata Jora
$^{\it \bf a}$~\footnote[1]{Email:
 rjora@theory.nipne.ro}}

\affiliation{$^{\bf \it a}$ National Institute of Physics and Nuclear Engineering PO Box MG-6, Bucharest-Magurele, Romania}

\date{\today}

\begin{abstract}
We make the assumption that the vacuum correlators of the gauge invariant kinetic term of the Higgs doublet are the same before and after the spontaneous symmetry breaking of the theory. Based on this we determine the mass of the standard model Higgs boson at $m_h \approx 125.07$ GeV by considering one loop and the most relevant two loop corrections. This result might suggest that there is a single Higgs boson doublet that contributes to the electroweak symmetry breaking.

\end{abstract}
\maketitle

\section{Introduction}
The standard model of elementary particles \cite{Glashow}-\cite{Hagen} has received its most important experimental confirmation with the discovery of the Higgs boson by the Atlas \cite{Atlas} and CMS \cite{CMS} experiments.
One of the most important theoretical issue associated with the standard model Higgs boson is that of naturalness which stems from the existence of significant quadratic corrections to the Higgs boson mass \cite{Veltman}. The standard model Higgs boson properties and higher loop corrections to the mass have been discussed in \cite{Isidori} and \cite{Martin}.  The bounds on the mass of the Higgs boson from the electroweak precision data were derived in \cite{Erler}. In order to predict the mass of the Higgs boson one needs additional assumptions or hypothesis as the mass of the Higgs boson cannot be extracted only from the standard model parameters.  In \cite{Jora} we determined the approximate mass of the Higgs boson from properties of the background gauge field. The correct mass for the Higgs boson was also obtained in \cite{Wetterich} from asymptotic safety of gravity, in \cite{H} from the three loop effective potential and in \cite{Carson} from 5D gauge unification.

In this work we rely on the properties of the standard model Lagrangian and partition function  before and after spontaneous symmetry breaking to make the assumption that the gauge invariant operators are conserved at the transition between the symmetric and spontaneously broken phases.  Based on this we calculate at one loop quantum correlators to which we add the most important two loop contributions to determine a mass of the Higgs boson very close to the central experimental value. This is done by relating the quadratic corrections to the gauge invariant kinetic term for the Higgs doublet for the symmetric and spontaneously broken Lagrangians.

In section II we review the standard model Lagrangian and present in detail the terms that are most relevant for the present work. In section III we compute the vacuum expectation values of the gauge invariant kinetic terms for the Higgs boson before and after spontaneous symmetry breaking. Section IV is dedicated to the results and their signification.

\section{The Lagrangian}

We consider the electroweak sector of the standard model given by the Lagrangian \cite{Cheng}:

\begin{eqnarray}
{\cal L}={\cal L}_1+{\cal L}_2+{\cal L}_3+{\cal L}_4,
\label{lagr332}
\end{eqnarray}
where,
\begin{eqnarray}
{\cal L}_1=-\frac{1}{4}F_{\mu\nu}^iF^{\mu\nu i}-\frac{1}{4}G^{\mu\nu}G_{\mu\nu},
\label{first44}
\end{eqnarray}
with,
\begin{eqnarray}
&&F^i_{\mu\nu}=\partial_{\mu}A_{\nu}^i-\partial_{\nu}A_{\mu}^i+g\epsilon^{ijk}A_{\mu}^jA_{\nu}^k
\nonumber\\
&&G_{\mu\nu}=\partial_{\mu}B_{\nu}-\partial_{\nu}B_{\mu}.
\label{nottr54}
\end{eqnarray}
Furthermore,
\begin{eqnarray}
{\cal L}_2=\bar{\Psi}i\gamma^{\mu}D_{\mu}\Psi.
\label{sec343}
\end{eqnarray}
Here we used a generic notations for all fermion terms  and for example:
\begin{eqnarray}
D_{\mu}\Psi=(\partial_{\mu}-ig\vec{T}A_{\mu}-ig'\frac{Y}{2}B_{\mu})\Psi.
\label{ferm878}
\end{eqnarray}
Next,
\begin{eqnarray}
&&{\cal L}_3=(D^{\mu}\Phi)^{\dagger}(D_{\mu}\Phi)-V(\Phi)
\nonumber\\
&&{\cal L}_4=f_e\bar{l}_L\Phi e_R+f_u\bar{q}_L\tilde{\Phi}u_R+f_d\bar{q}_L\Phi d_R+h.c.
\label{forth554}
\end{eqnarray}
Here $\Phi$ is the standard model Higgs doublet,
\begin{eqnarray}
\Phi=
\left(
\begin{array}{c}
\Phi^{+}\\
\Phi_0
\end{array}
\right),
\label{higgs443}
\end{eqnarray}
with,
\begin{eqnarray}
&&D_{\mu}\Phi=(\partial_{\mu}-\frac{i}{2}g\vec{\tau}\vec{A}_{\mu}-\frac{i}{2}g'B_{\mu})\Phi
\nonumber\\
&&V(\Phi)=-m_0^2\Phi^{\dagger}\Phi+\lambda(\Phi^{\dagger}\Phi)^2
\nonumber\\
&&\tilde{\Phi}=i\tau_2\Phi^*.
\label{res2121}
\end{eqnarray}

The potential in Eq. (\ref{res2121}) displays spontaneous symmetry breaking according to the structure:
\begin{eqnarray}
\Phi=\frac{1}{\sqrt{2}}
\left(
\begin{array}{c}
\Phi^{\prime+}\\
v+h+i\Phi_3,
\end{array}
\right)
\label{dec3443}
\end{eqnarray}
where $v=\frac{m_0^2}{\lambda}$ is the vacuum expectation value, $h$ is the Higgs and $\Phi_3$, $\Phi^{\prime\pm}$ are the Goldstone bosons. As a result of spontaneous symmetry breaking the mass of the Higgs boson will become $2m_0^2$ whereas the Goldstone bosons will remain massless.

We are mainly interested in the gauge kinetic term for the Higgs boson. For future use we give below the detailed expression for this term as a function of the mass eigenstates of the gauge fields after spontaneous symmetry breaking:
\begin{eqnarray}
&&(\partial^{\mu}\Phi)^{\dagger}(\partial_{\mu}\Phi)=
\partial^{\mu}\Phi^{-}\partial_{\mu}\Phi^++\partial^{\mu}\Phi_0^*\partial_{\mu}\Phi_0+
\nonumber\\
&&\partial^{\mu}\Phi^{-}[-\frac{i}{2}(bZ_{\mu}+cA_{\mu})\Phi^+-\frac{i}{\sqrt{2}}gW_{\mu}^+\Phi_0g]+
\nonumber\\
&&\partial_{\mu}\Phi^{+}[\frac{i}{2}(bZ^{\mu}+cA^{\mu})\Phi^-+\frac{i}{\sqrt{2}}gW^{\mu-}\Phi_0g]+
\nonumber\\
&&\partial^{\mu}\Phi_0^*[-\frac{i}{\sqrt{2}}gW_{\mu}^-\Phi^+-\frac{i}{2}aZ_{\mu}\Phi_0]+
\nonumber\\
&&\partial_{\mu}\Phi_0[\frac{i}{\sqrt{2}}gW^{\mu+}\Phi^-+\frac{i}{2}aZ^{\mu}\Phi_0^*]+
\nonumber\\
&&[\frac{1}{2}(bZ^{\mu}+cA^{\mu})\Phi^-+\frac{1}{\sqrt{2}}gW^{\mu-}\Phi_0][\frac{1}{2}(bZ_{\mu}+cA_{\mu})\Phi^++\frac{1}{\sqrt{2}}gW_{\mu}^+\Phi_0]+
\nonumber\\
&&[\frac{1}{\sqrt{2}}W^{\mu-}\Phi^++\frac{a}{2}Z^{\mu}\Phi_0][\frac{1}{\sqrt{2}}W_{\mu}^+\Phi^-+\frac{a}{2}Z_{\mu}\Phi_0].
\label{expr6657867}
\end{eqnarray}
Here we made the notations:
\begin{eqnarray}
&&a=\sqrt{g^2+g^{\prime 2}}
\nonumber\\
&&b=\frac{g^2-g^{\prime 2}}{\sqrt{g^2+g^{\prime 2}}}
\nonumber\\
&&c=\frac{2gg^{\prime}}{\sqrt{g^2+g^{\prime 2}}}=2e.
\label{nottr5466}
\end{eqnarray}
The gauge kinetic term for the Higgs boson after spontaneous symmetry breaking can be read off very easily by applying Eq. (\ref{dec3443}) to  Eq. (\ref{expr6657867}).

\section{Invariant operators}

A gauge invariant Lagrangian needs fixing and one can show that the corresponding partition function remains unchanged through this procedure up to an irrelevant proportionality factor. In this work we shall make the assumption that the vacuum correlators of the gauge invariant operators are exactly the same before and after spontaneous symmetry breaking. In other words the gauge invariant  quantum correlator calculated for the symmetric partition function that contains tachyonic states is equal to the corresponding quantity calculated for the partition function with spontaneously broken symmetry.   This assumption is justified by the fact that in the actual partition function the transition from one state to another can be simply expressed in a change of variable in the Higgs sector $\Phi_0=\frac{1}{\sqrt{2}}(v+h+i\Phi_3)$. In order for this equivalence of the quantum correlators to be valid it is necessary to keep, besides this change of variable, the partition function unchanged before and after spontaneous symmetry breaking. This implies that the gauge fixing of the Lagrangian must be done similarly and consistent in both cases without the introduction of additional terms in the Lagrangian. The most amenable gauge is the Landau gauge $\xi=0$ which will be used before and after spontaneous symmetry breaking with the amend that the gauge fixing function must be the same in both situations. This a slight depart from the $R_{\xi}$ gauges  for theories with spontaneous symmetry breaking where for example the gauge function for a generic case is if the type $G=\frac{1}{\sqrt{\xi}}(\partial^{\mu}A_{\mu}-\xi v\phi)$ where $\phi$ is the Goldstone boson associated to the gauge field $A_{\mu}$. Here we shall take simply $G=\frac{1}{\sqrt{\xi}}(\partial^{\mu}A_{\mu})$ for both cases.

In the standard model Lagrangian there are quite a few of gauge invariant operators but we are interested in those that contain the Higgs doublet and no constant terms.  The most relevant among these is the gauge kinetic term for the Higgs doublet. We shall equate the quadratic corrections at one loop before and after spontaneous symmetry breaking. Before spontaneous symmetry breaking this term reads:
\begin{eqnarray}
(D^{\mu}\Phi^{\dagger})(D_{\mu}\Phi)=\partial^{\mu}\Phi^-\partial_{\mu}\Phi^++\partial^{\mu}\Phi_0^*\partial_{\mu}\Phi_0+{\rm trilinear\,\,and\,\, quadrilinear\,\,terms}
\label{bef55464}
\end{eqnarray}
whereas after spontaneous symmetry breaking:
\begin{eqnarray}
&&(D^{\mu}\Phi^{\dagger})(D_{\mu}\Phi)=
\nonumber\\
&&\frac{1}{2}\partial^{\mu}h\partial_{\mu}h+\frac{1}{2}\partial^{\mu}\Phi^{\prime-}\partial_{\mu}\Phi^{\prime+}+\frac{1}{2}\partial^{\mu}\Phi_3\partial_{\mu}\Phi_3+
\nonumber\\
&&-m_Z^2Z^{\mu}\partial_{\mu}\Phi_3-i\frac{1}{\sqrt{2}}W^{\mu+}\partial_{\mu}\Phi^{\prime-}+i\frac{1}{\sqrt{2}}W_{\mu}^-\partial^{\mu}\Phi^{\prime+}+
\nonumber\\
&&{\rm trilinear\,\,and\,\,quadrilinear\,\,terms}.
\label{after42334}
\end{eqnarray}

Next we compute the quadratic corrections corresponding to Eq. (\ref{bef55464}):
\begin{eqnarray}
A=\langle\int d^4 x (D^{\mu}\Phi^{\dagger})(D_{\mu}\Phi)\rangle=\delta(0)\int_{|\vec{p}|\leq m_0} \frac{d^4p}{(2\pi)^4}i\frac{p^2}{p^2+m_0^2+i\epsilon}\frac{1}{2}\times 4,
\label{invf5454}
\end{eqnarray}
where the factor $4$ comes from the four scalars with the same mass in the Higgs doublet (see Appendix A for the choice of the tachyon propagator and for how the tachyon contribution is treated). The function $\delta(0)$ is in the momentum space and establishes the correct dimensionality of the integral.
Similarly after spontaneous symmetry breaking the same  quantum correlator yields:
\begin{eqnarray}
&&B=\langle(D^{\mu}\Phi^{\dagger})D_{\mu}\Phi)\rangle=
\nonumber\\
&&\delta(0)\int \frac{d^4p}{(2\pi)^4}[\frac{i}{2}\frac{p^2}{p^2-m_h^2}+\frac{3i}{2}\frac{p^2}{p^2}-i\frac{3}{2}m_Z^2\frac{1}{p^2-m_Z^2}-3i\frac{1}{p^2-m_W^2}].
\label{res554664}
\end{eqnarray}
The calculations are done as usual by rotating to the euclidean space. We use,
\begin{eqnarray}
\int \frac{d^4p}{(2\pi)^4}\frac{1}{p^2-m_X^2}=-\frac{i}{16\pi^2}[\Lambda^2-m_X^2\ln[\frac{\Lambda^2}{m_X^2}]],
\label{int43789}
\end{eqnarray}
to find  by equating the quadratic divergences before and after spontaneous symmetry breaking (see also Appendix A):
\begin{eqnarray}
-2m_0^2=-\frac{3}{2}m_Z^2-3m_W^2+\frac{m_h^2}{2}
\label{rel775645}
\end{eqnarray}
Then by assuming the bare masses $m_h^2=2m_0^2$ we determine:
\begin{eqnarray}
m_h^2=m_Z^2+2m_W^2.
\label{rqwew}
\end{eqnarray}
Eq. (\ref{rqwew}) leads, by introducing the physical masses for the gauge boson to an estimate $m_h\approx145.7$ GeV which is very large compared to the known physical value of the Higgs boson.

However this result is a gross approximation for several reasons. First the quadratic term that relates the Goldstone boson with the gauge boson contributes significantly at one loop. Second the two loop contributions especially those that involve the top quark and the tadpole diagrams give important quadratic corrections  at two loop which alter substantially the one loop result. Next one should consider also the one loop corrections to the bare mass in order to extract a suitable mass for the Higgs boson since these corrections contains quadratic divergences. With one amend that we will state later we shall consider all of the above arguments in our subsequent calculations.

We start by considering the quadratic corrections that come from the gauge boson Goldstone boson term. We illustrate this only for the $Z_{\mu}$ boson:
\begin{eqnarray}
&&\langle\int d^4x d^4 y (-m_Z\partial^{\mu}\Phi_3(x)Z_{\mu}(x))(-im_Z\partial_{\nu}\Phi_3(y)Z^{\nu}(y))\rangle=
\nonumber\\
&&\delta(0)\frac{3m_Z^2}{16\pi^2}[\Lambda^2-m_Z^2\ln[\frac{\Lambda^2}{m_Z^2}]]
\label{qrew67}
\end{eqnarray}
A similar contribution is obtained for the $W_{\mu}^{\pm}$ bosons  for which we shall write down only the result:
\begin{eqnarray}
&&\langle\int\int d^4x d^4 y 2(\frac{-i}{\sqrt{2}}m_W\partial^{\mu}\Phi^{\prime+}(x)W_{\mu}^-(x))(i\frac{i}{\sqrt{2}}m_W\partial_{\nu}\Phi^{\prime-}(y)W^{\nu}(y))\rangle=
\nonumber\\
&&\delta(0)\frac{6m_W^2}{16\pi^2}[\Lambda^2-m_W^2\ln[\frac{\Lambda^2}{m_W^2}]].
\label{gaugeb554664}
\end{eqnarray}.
If we add the contribution in Eqs. (\ref{qrew67}) and (\ref{gaugeb554664}) to the leading one loop quadratic corrections in Eq. (\ref{rel775645}) we obtain:
\begin{eqnarray}
&&-2m_0^2=-\frac{3}{2}m_Z^2-3m_W^2+3m_Z^2+6m_W^2+\frac{m_h^2}{2}
\nonumber\\
&&-2m_0^2=\frac{3}{2}m_Z^2+3m_W^2+\frac{m_h^2}{2},
\label{res97539}
\end{eqnarray}
which does not make sense as it is and clearly shows the need for the two loop corrections.

 In general a cut-off regularization procedure may be just for one loop corrections but not quite so for higher loops because the translation and gauge invariance are lost. However because we are interested in quadratic corrections we shall proceed further and consider only those Feynman diagrams that are well behaved under a cut-off procedure and neglect those more intricate that require the introduction of the Feynman parameters. Namely the diagrams that we will take into account are those diagrams that may be written as product of two loops or vacuum diagrams that contain two connected bubbles, be they of the tadpole type. Since the diagrams we consider contain product of two traces and those that we ignore contain one trace and mostly Goldstone bosons we estimate that these latter are strongly suppressed and at most may contribute to the final result by one percent.

 Below we shall briefly enumerate those two loop contributions that we shall take into account for which we are interested mainly in corrections of order $\Lambda^2$.
\vspace{0.5cm}

1) Tadpole diagram involving the top quark and the $Z_{\mu}$ boson:
\begin{eqnarray}
&&I_1=\langle\int d^4x d^4 y m_Z^2Z^{\mu}(x)Z_{\mu}(x)\frac{h(x)}{v}(-i)\frac{m_t}{v}\bar{t}(y)t(y)h(y)\rangle=
\nonumber\\
&&-36\delta(0)\frac{m_Z^2m_t^2}{m_h^2v^2}\frac{1}{256\pi^4}(\Lambda^2-m_Z^2\ln[\frac{\Lambda^2}{m_Z^2}])(\Lambda^2-m_t^2\ln[\frac{\Lambda^2}{m_t^2}]).
\label{top6574}
\end{eqnarray}
\vspace{0.5cm}
2) Tadpole diagram involving the top quark and the $W^{\pm}_{\mu}$ boson:
\begin{eqnarray}
&&I_2=\langle\int d^4x d^4 y2 m_W^2W^{+\mu}(x)W_{\mu}^-(x)\frac{h(x)}{v}(-i)\frac{m_t}{v}\bar{t}(y)t(y)h(y)\rangle=
\nonumber\\
&&-72\delta(0)\frac{m_W^2m_t^2}{m_h^2v^2}(\Lambda^2-m_W^2\ln[\frac{\Lambda^2}{m_W^2}])(\Lambda^2-m_t^2\ln[\frac{\Lambda^2}{m_t^2}]).
\label{top657434}
\end{eqnarray}
\vspace{0.5cm}
3) Tadpole diagrams that contains four $Z_{\mu}$ bosons:
\begin{eqnarray}
&&I_3=\langle\int d^4 x d^4 y m_Z^2Z^{\mu}(x)Z_{\mu}(x)\frac{h(x)}{v}im_Z^2Z^{\mu}(y)Z_{\mu}(y)\frac{h(y)}{v}\rangle=
\nonumber\\
&&9\delta(0)\frac{m_Z^4}{m_h^2v^2}\frac{1}{256\pi^4}(\Lambda^2-m_Z^2\ln[\frac{\Lambda^2}{m_Z^2}])(\Lambda^2-m_Z^2\ln[\frac{\Lambda^2}{m_Z^2}]).
\label{third546788}
\end{eqnarray}
\vspace{0.5cm}
4) Tadpole diagram that contains four $W_{\mu}^{\pm}$ bosons:
\begin{eqnarray}
&&I_4=\langle\int d^4 x d^4 y 4 m_W^2W^{\mu+}(x)W_{\mu}^-(x)\frac{h(x)}{v}im_W^2W^{\mu+}(y)W_{\mu}^-(y)\frac{h(y)}{v}\rangle=
\nonumber\\
&&36\delta(0)\frac{m_W^4}{m_h^2v^2}\frac{1}{256\pi^4}(\Lambda^2-m_W^2\ln[\frac{\Lambda^2}{m_W^2}])(\Lambda^2-m_W^2\ln[\frac{\Lambda^2}{m_W^2}]).
\label{third5467889}
\end{eqnarray}
\vspace{0.5cm}
5) Tadpole diagram that contains two $Z_{\mu}$ bosons and two $W_{\mu}^{\pm}$ bosons:
\begin{eqnarray}
&&I_5=\langle\int d^4 x d^4 y 4 m_W^2W^{\mu+}(x)W_{\mu-}(x)\frac{h(x)}{v}im_z^2Z^{\mu}(y)Z_{\mu}(y)\frac{h(y)}{v}\rangle=
\nonumber\\
&&36\delta(0)\frac{m_W^2m_Z^2}{m_h^2v^2}\frac{1}{256\pi^4}(\Lambda^2-m_W^2\ln[\frac{\Lambda^2}{m_W^2}])(\Lambda^2-m_Z^2\ln[\frac{\Lambda^2}{m_Z^2}]).
\label{third54678854}
\end{eqnarray}

\vspace{0.5cm}

6) Tadpole diagrams involving four Higgs bosons and two $Z_{\mu}$.
\begin{eqnarray}
&&I_6=\langle\int d^4x d^4 y m_Z^2Z^{\mu}(x)Z_{\mu}(x)\frac{h(x)}{v}\lambda v h^3(y)\rangle=
\nonumber\\
&&9\delta(0)\frac{m_Z^2}{2v^2}\frac{1}{256\pi^4}(\Lambda^2-m_h^2\ln[\frac{\Lambda^2}{m_h^2}])(\Lambda^2-m_Z^2\ln[\frac{\Lambda^2}{m_Z^2}]).
\label{rs1212}
\end{eqnarray}
\vspace{0.5cm}

7)  Tadpole diagram involving four Higgs bosons and two $W_{\mu}^{\pm}$ bosons:
\begin{eqnarray}
&&I_7=\langle\int d^4x d^4 y 2m_W^2W^{\mu+}(x)W_{\mu}^-(x)\frac{h(x)}{v}\lambda v h^3(y)\rangle=
\nonumber\\
&&18\delta(0)\frac{m_W^2}{2v^2}\frac{1}{256\pi^4}(\Lambda^2-m_h^2\ln[\frac{\Lambda^2}{m_h^2}])(\Lambda^2-m_W^2\ln[\frac{\Lambda^2}{m_W^2}]).
\label{rs121342}
\end{eqnarray}

\vspace{0.5cm}

8) Diagram with two Higgs bosons and two $Z_{\mu}$:
\begin{eqnarray}
&&I_8=\langle\int d^4 x \frac{m_Z^2}{2}Z^{\mu}(x)Z_{\mu}(x)\frac{h(x)^2}{v^2}\rangle=
\nonumber\\
&&-\frac{3}{2}\delta(0)\frac{m_Z^2}{v^2}\frac{1}{256\pi^4}(\Lambda^2-m_Z^2\ln[\frac{\Lambda^2}{m_Z^2}])(\Lambda^2-m_h^2\ln[\frac{\Lambda^2}{m_h^2}]).
\label{eight665}
\end{eqnarray}
\vspace{0.5cm}

9)  Diagram with two Higgs bosons and two $W_{\mu}^{\pm}$:

\begin{eqnarray}
&&I_9=\langle\int d^4 x\frac{m_W^2}{v^2} W^{\mu+}(x)W_{\mu}^-(x)\frac{h(x)^2}{v^2}\rangle=
\nonumber\\
&&-3\delta(0)\frac{m_W^2}{v^2}\frac{1}{256\pi^4}(\Lambda^2-m_W^2\ln[\frac{\Lambda^2}{m_W^2}])(\Lambda^2-m_h^2\ln[\frac{\Lambda^2}{m_h^2}]).
\label{eight665}
\end{eqnarray}
\vspace{0.5cm}

10)   Diagrams with two $Z_{\mu}$ and Goldstone bosons after spontaneous symmetry breaking:
\begin{eqnarray}
&&I_{10}+I_{11}=\langle\int d^4 x \frac{m_Z^2}{2}Z^{\mu}(x)Z_{\mu}(x)\frac{\Phi_3^2(x)}{v^2}+\int d^4 x \frac{m_Z^2}{2}\frac{b^2}{a^2}Z^{\mu}(x)Z_{\mu}(x)\frac{\Phi^{\prime+}\Phi^{\prime-}}{v^2}\rangle=
\nonumber\\
&&-\frac{3}{2}\delta(0)\frac{m_Z^2}{v^2}\frac{1}{256\pi^4}(\Lambda^2-m_Z^2\ln[\frac{\Lambda^2}{m_Z^2}])\Lambda^2-3\frac{m_Z^2}{v^2}\frac{b^2}{a^2}\frac{1}{256\pi^4}(\Lambda^2-m_Z^2\ln[\frac{\Lambda^2}{m_Z^2}])\Lambda^2.
\label{goldst443}
\end{eqnarray}
\vspace{0.5cm}

11) Diagrams with two $W_{\mu}^{\pm}$ and Goldstone bosons after symmetry breaking:
\begin{eqnarray}
&&I_{12}+I_{13}=\langle\int d^4 x \frac{m_W^2}{v^2}W^{\mu+}(x)W_{\mu}^-(x)\frac{\Phi_3^2}{v^2}+\int d^4 x \frac{m_W^2}{v^2}W^{\mu+}(x)W_{\mu}^-(x)\frac{\Phi^{\prime+}\Phi^{\prime-}}{v^2}\rangle=
\nonumber\\
&&-3\delta(0)\frac{m_W^2}{v^2}\frac{1}{256\pi^4}(\Lambda^2-m_W^2\ln[\frac{\Lambda^2}{m_W^2}])\Lambda^2-6\frac{m_W^2}{v^2}\frac{1}{256\pi^4}(\Lambda^2-m_W^2\ln[\frac{\Lambda^2}{m_W^2}])\Lambda^2.
\label{goldstw4435}
\end{eqnarray}

\vspace{0.5cm}
12) Diagrams with two $Z_{\mu}$ and the four Higgses before spontaneous symmetry breaking:
\begin{eqnarray}
&&I_{14}+I_{15}=\langle\int d^4 x \frac{m_Z^2}{v^2}Z^{\mu}(x)Z_{\mu}(x)\frac{\Phi_0^2(x)}{v^2}+\int d^4 x \frac{m_Z^2}{v^2}\frac{b^2}{a^2}Z^{\mu}(x)Z_{\mu}(x)\frac{\Phi^{+}\Phi^{-}}{v^2}\rangle=
\nonumber\\
&&-3\delta(0)\frac{m_Z^2}{v^2}\frac{1}{256\pi^4}(\Lambda^2+m_0^2\ln[\frac{\Lambda^2}{m_0^2}])\Lambda^2-3\frac{m_Z^2}{v^2}\frac{b^2}{a^2}\frac{1}{256\pi^4}(\Lambda^2+m_0^2\ln[\frac{\Lambda^2}{m_0^2}])\Lambda^2.
\label{goldst443222}
\end{eqnarray}
\vspace{0.5cm}

13)  Diagrams with two $W_{\mu}^{\pm}$ and the four Higgses before spontaneous symmetry breaking:
\begin{eqnarray}
&&I_{16}=\langle\int d^4 x 2\frac{m_W^2}{v^2}W^{\mu+}(x)W_{\mu}^-(x)\frac{\Phi_0^2}{v^2}+\int d^4 x 2\frac{m_W^2}{v^2}W^{\mu+}(x)W_{\mu}^-(x)\frac{\Phi^{\prime+}\Phi^{\prime-}}{v^2}\rangle=
\nonumber\\
&&-3\delta(0)\frac{m_W^2}{v^2}\frac{1}{256\pi^4}4(\Lambda^2+m_0^2\ln[\frac{\Lambda^2}{m_0^2}])\Lambda^2.
\label{whiggs435664}
\end{eqnarray}

Before going further we need to make a remark. We considered only the bubble diagrams that involve product of integrals and ignored the most intricate ones that involve the Feynman parameters. The integrals we neglected are associated to different contractions to the fields in the points 3) and 4) above. First we observe that they contain one single trace over gauge degrees of freedom that amount to a factor of 3 (whereas the integrals in 3) and 4) contain a product of two traces that amount to 9). However we need to show that the actual integral given below in the euclidian space (we consider the case 3) here):
\begin{eqnarray}
A_1=3\int \frac{d^4p_1}{(2\pi)^4}\frac{d^4p_2}{(2\pi)^4}\frac{1}{(p_1-p_2)^2+m_h^2}\frac{1}{p_1^2+m_Z^2}\frac{1}{p_2^2+m_Z^2},
\label{firs5463int}
\end{eqnarray}
is reasonably smaller than the similar integral that is actually calculated in Eq. (\ref{third546788}):
\begin{eqnarray}
A_2=9\int \frac{d^4p_1}{(2\pi)^4}\frac{d^4p_2}{(2\pi)^4}\frac{1}{m_h^2}\frac{1}{p_1^2+m_Z^2}\frac{1}{p_2^2+m_Z^2}.
\label{sec53828565}
\end{eqnarray}
In order to show that we will use the H${\rm \ddot{o}}$lder inequality which states that for two functions real or complex valued $f(x)$ and $g(x)$ on a volume $V\subset R^n$ with Lebesgue measure  the following relation holds:
\begin{eqnarray}
\int  dx |f(x)g(x)|\leq (\int dx |f(x)|^p)^{1/p} (\int dx |g(x)|^q)^{1/q}
\label{res647586}
\end{eqnarray}
where $p$ and $q$ are two real numbers situated in the interval $(1,\infty)$ and that satisfy the relation $\frac{1}{p}+\frac{1}{q}=1$. We will choose $q=n$ and $p=\frac{n}{n-1}$  and in the end take the limit $n\rightarrow\infty$. This yields according to the H${\rm \ddot{o}}$lder inequality:
\begin{eqnarray}
&&A_1\leq (\int \frac{d^4p_1}{(2\pi)^4}\frac{d^4p_2}{(2\pi)^4}(\frac{1}{(p_1-p_2)^2+m_h^2})^{q})^{1/q}(\int \frac{d^4p_3}{(2\pi)^4}\frac{d^4p_4}{(2\pi)^4}\frac{1}{p_3^2+m_Z^2}\frac{1}{p_4^2+m_Z^2})^{p})^{1/p}=
\nonumber\\
&&3m_h^2\frac{A_2}{9}\lim_{n\rightarrow\infty}\Bigg[\frac{\Lambda^4}{512\pi^4}((m_h^2)^{2-n}-(m_h^2+\Lambda^2)^{1-n}(m_h^2+(n-1)\Lambda^2))\frac{1}{2-3n+n^2}\Bigg]^{1/n}=\frac{A_2}{3}
\label{res7536489}
\end{eqnarray}
Since $\frac{1}{3}$ is considered reasonably small in quantum field theories our approximation is thus justified.

In what follows we will ignore one loop corrections to the gauge boson and fermion masses as they are significantly smaller than the Higgs corrections to the mass and replace the bare masses with the physical ones. Alternatively one may assume that in the initial set up one considers the gauge invariant operator with the physical masses and  thus it is not necessary to introduce in the equations of interest the corresponding correction terms.
However we will consider corrections to the Higgs boson mass because they are important but apply them at the level of the mass parameter in the Lagrangian rather than in the propagator such that when appropriate we substitute the bare mass by the physical one. Note that this procedure ensures the presence of the Higgs physical mass in the propagator.
The one loop corrected Higgs boson mass is given by:
\begin{eqnarray}
m_h^2=2m_0^2+\Sigma(m_h^2),
\label{res55342}
\end{eqnarray}
where \cite{Ma},
\begin{eqnarray}
&&\Sigma(m_h^2)=\frac{3}{8\pi^2v^2}[4m_t^2-m_h^2-2m_W^2-m_Z^2]\Lambda^2-
\nonumber\\
&&\frac{3}{16\pi^2}\frac{m_h^2}{v^2}[m_h^2\ln[\frac{\Lambda^2}{m_h^2}]-2m_W^2\ln[\frac{\Lambda^2}{m_W^2}]-m_Z^2\ln[\frac{\Lambda^2}{m_Z^2}]+2m_t^2\ln[\frac{\Lambda^2}{m_t^2}]].
\label{res66453}
\end{eqnarray}
For simplicity we shall denote:
\begin{eqnarray}
\Sigma(m_h^2)=x\Lambda^2+y\ln[\Lambda^2]+z,
\label{not775657}
\end{eqnarray}
where $x$ is the coefficient of the quadratic divergent term, $y$ that of the logarithmic one and $z$ is the constant contribution. Note that whenever we use the logarithm of a quantity with mass dimension $m^2$  expressed in $GeV^2$ the division of the argument of the logarithm by $1$ $GeV^2$ is implicitly assumed.

Now we add all two loop contributions in the Eqs. (\ref{top6574})-(\ref{whiggs435664}) to the one loop contribution in Eq. (\ref{res97539}) and apply Eq. (\ref{res55342}) to obtain as a result of equating the gauge Higgs kinetic term before and after spontaneous symmetry breaking:
\begin{eqnarray}
&&\frac{3}{2}(m_h^2-z)-x(m_h^2-z)\ln[2]+x(\frac{m_h^2}{2}-z)\ln[m_h^2]=
\nonumber\\
&&=-3m_W^2-\frac{3}{2}m_Z^2-\sum_{j=1}^{16}(16\pi^2)\frac{1}{\delta(0)}(I_j)_{\Lambda^2},
\label{finalres66453}
\end{eqnarray}
where $(I_j)_{\Lambda^2}$ denotes the coefficient of the quadratic divergence in the integral $I_j$.

\section{Discussion and conclusions}
In this work we made the hypothesis that the quantum correlators associated to the space time integral of gauge invariant kinetic terms of the standard model Higgs boson doublet for the symmetric Lagrangian (partition function) and spontaneously broken Lagrangian (partition function) are equal. This assumption is based on the fact that if one uses the same gauge fixing functions for the Lagrangian before and after spontaneous symmetry breaking the only change that occurs in the Lagrangian and partition function at the symmetry breaking point is a change of variable by a constant shift of the neutral component of the Higgs doublet.

 We computed the corresponding quantities and their quadratic contribution at one loop with corrections from the most relevant diagrams at two loop.

 \begin{figure}
\begin{center}
\epsfxsize = 10cm
\epsfbox{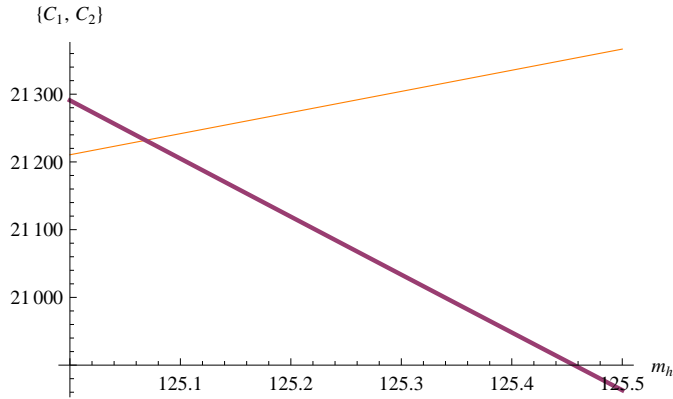}
\end{center}
\caption[]{%
Plot of the  quantities $C_1$ ($GeV^2$) and $C_2$ ($GeV^2$) as a function of the Higgs boson mass $m_h$ ($GeV$).
}
\label{masses}
\end{figure}

In Fig.1 we plotted $C_1$ and $C_2$ where:
\begin{eqnarray}
&&C_1=\frac{3}{2}(m_h^2-z)-x(m_h^2-z)\ln[2]+x(\frac{m_h^2}{2}-z)\ln[m_h^2]
\nonumber\\
&&C_2=-3m_W^2-\frac{3}{2}m_Z^2-\sum_{j=1}^{16}(16\pi^2)(I_j)_{\Lambda^2},
\label{restfinal657}
\end{eqnarray}
to find that the two curves intersect for a mass of the Higgs boson $m_h\approx 125.07$ GeV. This result is strikingly close to the experimental mass (\cite{PDG}) of the Higgs boson $m_{hexp}=125.09\pm 0.24$ GeV. We estimate  the other two loop diagrams contribution or corrections to the masses of the other particles involved besides the Higgs  for at most 1 percent of the value obtained in this work.  Our finding suggest that our initial assumption may be well justified.

If a beyond the standard model theory is employed instead then our hypothesis should apply to each gauge invariant kinetic operator for the scalars involved and the results should be greatly altered due to different possible combinations of vacuum expectation values and masses. Consequently it is very possible that the electroweak breaking sector of the standard model might contain at least in first order as an effective theory  a single Higgs doublet with the well known vacuum expectation value and Goldstone bosons.

\begin{appendix}
\section{}
The tachyon propagator has been studied in the literature with different results \cite{Feinberg}-\cite{Rocca}. In general it is expressed in terms of the advances and retarded Green functions with the sphere $|\vec{k}|^2<m^2$ removed from the regular integral (where $m^2>0$ is the tachyonic mass). Here however we shall consider the point of view adopted in Perepelitsa \cite{Perepelitsa} where an extra term is added in the tachyon Lagrangian leading to a completely equivalent one. The corresponding propagator is then (if $\Phi(x)$ is the tachyon field):
\begin{eqnarray}
\langle 0|T\Phi(x)\Phi(y)|0\rangle=\int_{|\vec{k}|\geq m}\frac{d^4k}{(2\pi)^4}\frac{i}{k^2+m^2+i\epsilon}\exp[-ik(x-y)].
\label{tprop9786}
\end{eqnarray}
In our calculations we need only the quantity:
\begin{eqnarray}
K=\langle 0|T\Phi(x)\Phi(x)|0\rangle=\int_{|\vec{k}|\geq m}\frac{d^4k}{(2\pi)^4}\frac{i}{k^2+m^2+i\epsilon}.
\label{t3452678}
\end{eqnarray}
The presence of the $i\epsilon$ term in the denominator together with the limits of integration makes very convenable the conversion to the euclidian space which is done in a standard way:
\begin{eqnarray}
K=\frac{1}{(2\pi)^4}\int_{|\vec{k}|\geq m}dk_E^0 dk_E ^1 dk_E^2 dk_E^3\frac{1}{(k^{0}_E)^2+|\vec{k}_E|^2-m^2}.
\label{euclid54674}
\end{eqnarray}
We consider the three dimensional sphere in the space coordinates to get:
\begin{eqnarray}
K=\frac{1}{(2\pi^4)}4\pi \int_{k^R_E\geq m}dk_E^0 d k_E^R (k_E^{R})^2\frac{1}{(k^{0}_E)^2+(k^{R}_E)^2-m^2}.
\label{res32789}
\end{eqnarray}
In order to solve this integral we introduce the polar coordinates in the momenta space:
\begin{eqnarray}
&&k^0_E=p\cos[\phi]
\nonumber\\
&&k^R_E=p\sin[\phi],
\label{pol8768}
\end{eqnarray}
with the angle $\phi$ in the interval $[0,\pi]$ and with the constraint $p\leq \frac{m}{\sin[\phi]}$. This leads to:
\begin{eqnarray}
K=\frac{1}{4\pi^3}\int_0^{\pi}d\phi \int_{\frac{m}{\sin[\phi]}}^{\Lambda} d p   p^3(\sin[\phi])^2\frac{1}{p^2-m^2}.
\label{finalint65748}
\end{eqnarray}
This integral can be solved exactly (one can add an infrared regulator) and has the result:
\begin{eqnarray}
K=\frac{1}{16\pi^2}[\Lambda^2+m^2\ln[\frac{\Lambda^2}{m^2}]]
\label{res537899}
\end{eqnarray}
and this is the expression employed in this  paper for the tachyon contribution.

\end{appendix}

\section*{Acknowledgments} \vskip -.5cm

The work of R. J. was supported by a grant of the Ministry of National Education, CNCS-UEFISCDI, project number PN-II-ID-PCE-2012-4-0078.

\end{document}